\documentclass[12pt,a4paper]{iopart}
\usepackage{iopams}
\usepackage{amssymb}
\usepackage{harvard}
\usepackage{graphicx}
\usepackage{subfigure}
\usepackage{epstopdf}
\usepackage{float}
\usepackage{rotating,booktabs}
\usepackage{multirow}
\usepackage[colorinlistoftodos]{todonotes}
\usepackage{hyperref}
\usepackage{booktabs}
\usepackage{comment}
\usepackage{xcolor}
\usepackage[normalem]{ulem} %RTN: for strike through

\hypersetup{
 breaklinks=true,
 pdftitle={title},    	% title
pdfauthor={Tom Bruijnen},      
 pdfsubject={title},   % subject of the document
 pdfkeywords={key1, key2},
 pdfcreator={tbruijnen},   % creator of the document
 pdfproducer={UMC Utrecht}}

\begin{document}

\title[]{Technical feasibility of Magnetic Resonance Fingerprinting on a 1.5T MRI-Linac}

\author{T. Bruijnen$^{1,2}$, O. van der Heide$^{2}$, M.P.W. Intven$^{1}$, S. Mook$^{1}$, J.J.W. Lagendijk$^{1}$, C.A.T. van den Berg$^{1,2}$ and R.H.N. Tijssen$^{1,3}$}

\address{$^1$ Department of Radiation Oncology, University Medical Center Utrecht, Utrecht, the Netherlands}
\address{$^2$ Computational Imaging Group for MRI diagnostics and therapy, Centre for Image Sciences, University Medical Center Utrecht, Utrecht, the Netherlands}
\address{$^3$  Department of Radiation Oncology, Catharina Hospital, Eindhoven, the Netherlands}

\begin{indented}
\item[]\today
\end{indented}

%% Abstract word: 223/250

\begin{abstract} 
Hybrid MRI-linac (\textbf{MRL}) systems enable daily multiparametric quantitative MRI to assess tumor response to radiotherapy. Magnetic Resonance Fingerprinting (\textbf{MRF}) may provide time efficient means of rapid multiparametric quantitative MRI. The accuracy of MRF, however, relies on adequate control over system imperfections, such as eddy currents and $B_1^+$, which are different and not as well established on MRL systems compared to diagnostic systems. In this study we investigate the technical feasibility of gradient spoiled 2D MRF on a 1.5T MRL. We show with phantom experiments that the MRL generates reliable MRF signals that are temporally stable during the day and have good agreement with spin-echo reference measurements. Subsequent \textit{in-vivo} MRF scans in healthy volunteers and a patient with a colorectal liver metastasis showed good image quality, where the quantitative values of selected organs corresponded with the values reported in literature. Therefore we conclude that gradient spoiled 2D MRF is feasible on a 1.5T MRL with similar performance as on a diagnostic system. The precision and accuracy of the parametric maps are sufficient for further investigation of the clinical utility of MRF for online quantitatively MRI-guided radiotherapy.

\end{abstract}
% \section*{key}

\submitto{Phys. Med. Biol. as a technical note}

\newpage
\section{Introduction}
One of the promises of magnetic resonance guided radiation therapy on hybrid MRI-linac (MRL) systems \cite{Lagendijk2008,Mutic2014,Fallone2014,Keall2014} is the ability to assess tumor response on a daily basis. The daily response is currently assessed using anatomical imaging, but could be replaced with precise quantitative imaging techniques  \cite{VanderHeide2018,Hall2019,Kooreman2019}. Traditional quantitative imaging techniques based on steady-state methods, such as variable flip angle ($T_1$-mapping) \cite{Fram1987} or multi-echo spin echo ($T_2$-mapping) \cite{Meiboom1958}, however, require long acquisition times. The long scan times pose a considerable practical challenge as the on table time is almost entirely filled with anatomical imaging (i.e., high-resolution 3D anatomical imaging for daily plan adaptation and fast real-time imaging for tumor tracking). Typical MRL treatment fractions have at most a couple of minutes of free imaging time available such that it does not interfere with the clinical workflow \cite{Raaymakers2017}(Fig.1-A). Therefore, the dual requirement of both fast and precise measurements mandates a sequence with a high precision per unit of time, i.e. quantification efficiency, for a practical implementation of online quantitative MRI-guided radiotherapy. Recently, transient-state-based quantitative imaging methods have been proposed to considerably improve this quantification efficiency. Magnetic resonance fingerprinting (MRF) \cite{Ma2013,Jiang2015,Jiang2017} is such a transient-state method that enables rapid multiparametric imaging and therefore could be the ideal tool for therapy monitoring on the MRL. \newline   

\begin{figure}
\begin{center}
   \includegraphics[width=16cm]{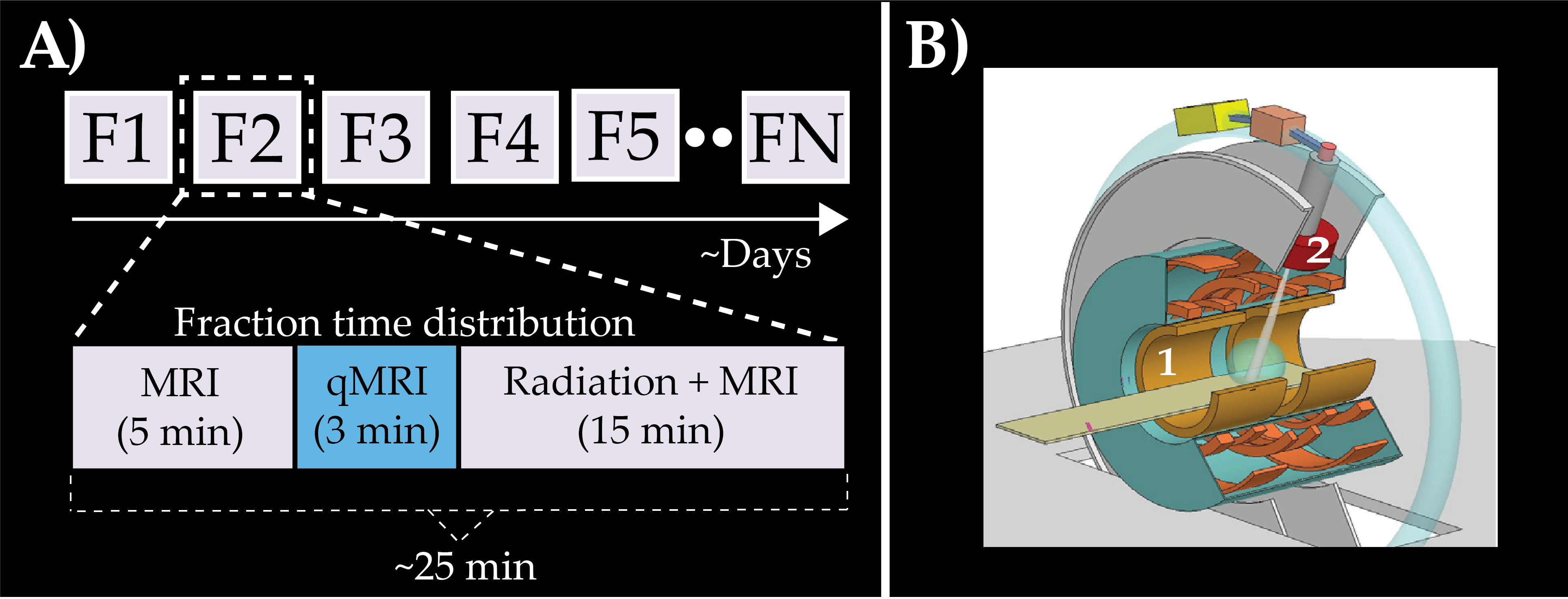}
 \vspace{-5pt}
\caption[]{\label{fig:fig1} \textbf{Schematic of a MRL radiotherapy treatment schedule.} \textbf{A)} Example of a fractionation scheme where a patient is treated over multiple radiation fractions (F) distributed across multiple days. During each fraction, MR images are required for multiple purposes. During the first $\approx$ 5 minutes anatomical images are acquired for radiotherapy treatment planning. The treatment planning takes around 5 min (depending on plan complexity), which can be used for quantitative imaging. Finally the irradiation is started and requires continuous anatomical imaging for motion management. \textbf{B)} Impression of the MRL (Unity, Elekta) with the split gradient coil shown in yellow and the linear accelerator gun shown in red, as indicated with number 1 and 2 respectively. }
\end{center}
\end{figure}

\noindent Unlike steady-state methods, which aim to produce a constant MR signal over time, MRF deliberately creates a fluctuating signal (fingerprint) over the course of the acquisition, which is matched on a per voxel basis to a precomputed dictionary of signal responses during image reconstruction. The dictionary is populated with simulated responses for all possible tissue types, in terms of $T_1, T_2$ and proton density, to the imposed transient-state MR sequence. Realistic simulations of these MR sequences are essential to accurately quantify the tissue properties \cite{Ma2017,Asslander2017}. 
Therefore, the simulation of the MR sequences requires adequate control over system imperfections such as eddy currents and heterogeneous magnetic fields ($B_{0}$ and $B_1^+$). These system imperfections are well characterised and controlled for diagnostic MR systems and MRF has been applied in multiple clinical studies \cite{Badve2017,Rieger2018,Liao2018,Ma2019,Yu2017,Chen2016,Chen2019,Cavallo2019}. However, these system imperfections are not yet accurately mapped in MRL systems. 
The 1.5T MRL system used in our institution (Unity, Elekta, Crawley, UK) differs from diagnostic systems in the split gradient and split magnet coil design, the radiolucent 2x4 channel receive coil and a paramagnetic rotating gantry that holds all the beam generating components (Fig.1-B). These hardware modifications have an impact on the system imperfections, such as reduced signal-to-noise ratio \cite{Hoogcarspel2018,Zijlema2019}, reduced uniformity of the static magnetic fields ($B_1^+$ and $B_{0}$) \cite{Crijns2014,Jackson2019}, reduced spatial region of gradient linearity \cite{Tijssen2019} and different behavior of the eddy currents \cite{Bruijnen2018}.
The impact of these system imperfections on the accuracy and precision of MRF parameter quantification is unknown. Therefore, an experimental study on the precision and accuracy of MRF is crucial for the potential application of daily quantitative tumor response monitoring on a 1.5T MRL.  \newline

\noindent In this work we investigate the technical feasibility of 2D MRF in phantoms and \textit{in-vivo} on a 1.5T MRL. We assess the accuracy, precision and temporal stability of the parameter quantification in a phantom. In addition, we showcase typical image quality of the parameter maps in comparison with clinically used qualitative scans in volunteers and patients. \newline 

\newpage
\section{Materials and methods}
\subsection{MRF pulse sequence and reconstruction method}
A 2D gradient spoiled MRF pulse sequence was implemented on a 1.5T Unity MR-linac equiped with a 2x4 channel radiation translucent receive array. Imaging data were acquired using the MRF sequence described by Jiang et al.\cite{Jiang2015}, which consists of an adiabatic inversion pulse and a sinusoidal flip angle train. One radial line was acquired per time-point \cite{Cloos2016} and subsequent readouts were azimuthally incremented using the tiny golden angle to minimize eddy current effects \cite{Wundrak2015,Bruijnen2019} (Fig.2). K-space data and k-space trajectory were corrected using the zeroth and first order gradient impulse response functions \cite{Vannesjo2013,Bruijnen2018}. Tissue fingerprints were simulated with extended phase graphs \cite{Weigel2015} with $T_1 \in [100:20:3000]$, $T_2 \in [20:10:1000]$ and inclusion of the slice profile \cite{Ma2017}. All data were used to estimate the coil sensitivities using ESPIRiT \cite{Uecker}. MRF k-space data were reconstructed into singular value images with low rank inversion \cite{Asslander2018} using the BART toolbox \cite{Uecker}. The singular value images were subsequently matched with the dictionary to reconstruct the parametric maps. The code to perform the image reconstruction code and one MRF dataset are available on  https://github.com/tombruijnen/mrf-mrl . 

\begin{figure}[H]
\begin{center}
   \includegraphics[width=16cm]{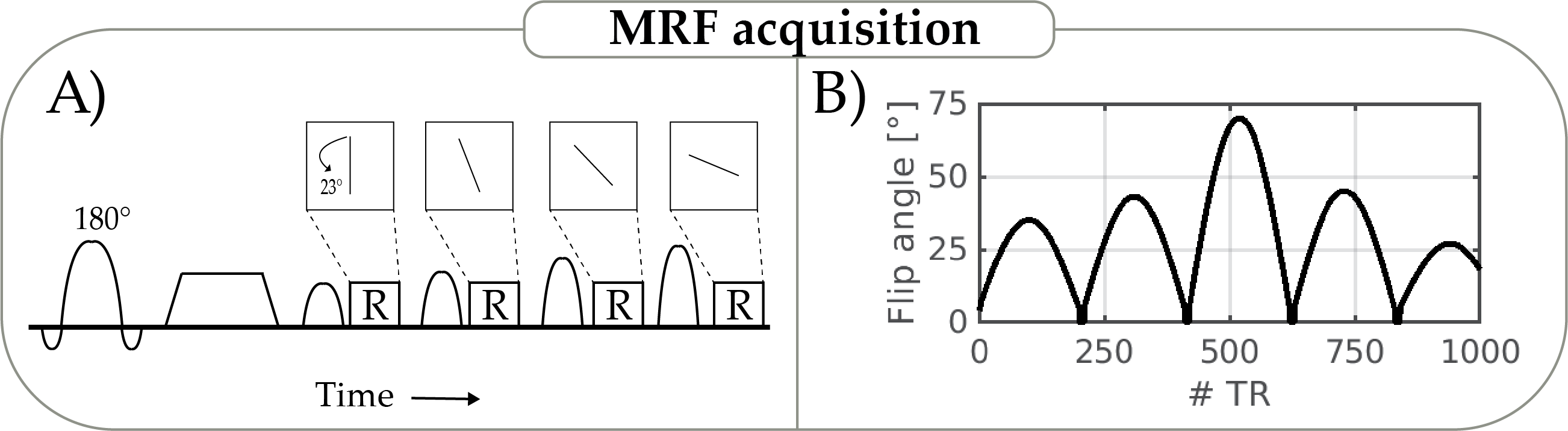}
 \vspace{-5pt}
\caption[]{\label{fig:fig2}\textbf{MRF acquisition overview. A)} The MRF scan consists of an adiabatic inversion pulse followed 1000 radial readouts. The radial readouts are rotated with the tiny golden angle for each repetition time. \textbf{B)} The flip angle train used for all experiments. Note that the four flip angles shown in panel A) reflect the first four flip angles in the train. }\vspace{-5pt}
\end{center}
\end{figure}

\subsection{Phantom studies}
MRF data were acquired in a 2D transverse slice of 14 gadolinium-doped gel tubes (TO5, Eurospin II test system, Scotland). Relevant sequence parameters for all scans are shown in Table.1. One fully sampled dataset was acquired, which consists of 276 repeated measurements of the MRF flip angle train, where for each measurement the azimuthal angle of the first spoke was rotated with the golden angle ($111.2^{\circ}$). The measurements had a 10 second interval between repetitions to allow for full spin relaxation. Three and sixteen hours later MRF measurements were repeated with two minutes intervals for 30 minutes. Note that the fully sampled MRF scans were reconstructed with the maximum correlation method \cite{Ma2013}. In total these scans provide 306 MRF measurements, which are used to estimate the precision, temporal stability and accuracy of the parameter quantification. The precision was quantified by calculating the standard deviation of parameter values within a tube. The temporal stability was quantified by calculating the standard deviation of the mean value within the tube over the repeated measurements. The accuracy was quantified by calculating the mean value within a tube and comparing it to reference measurements. The reference measurements were acquired using two separate inversion recovery ($T_1$) and spin-echo ($T_2$) scans. The reference data were acquired with single echo spin-echo measurements with: voxel size = 3x3x10 $mm^3$ scan time = 60 min, repetition time = 10 s and 10 inversion times $\in$ [100:3000] ms or 10 echo times $\in$ [20:500] ms. 

\subsection{\textit{In-vivo} studies}
This  study  was  approved  by  the institutional  review  board of the UMC Utrecht (Medisch Ethische Toetsingscommissie Utrecht (METC), ID:17-010, "MRI protocol development for MR-linac") and informed consent was obtained from all the participants. MRF data were acquired in the brain and upper abdomen of a healthy volunteer. One patient with a recurrent colorectal liver metastasis, after hepatic surgery, was scanned using the described MRF sequence with the addition of an custom developed abdominal compression corset to reduce motion artefacts \cite{Heerkens2017}. The complete MRI protocol consisted of multiple 2D MRF scans and qualitative $T_{1}$ and $T_{2}$-\textit{w} scans derived from the clinical protocol. The MRF scans in the upper abdomen were scanned in breathhold for the volunteer and in free-breathing for the patient. Regions of interest were manually selected on specific organs to compute the mean values, which were compared to literature values \cite{DeBazelaire2004,Deoni2005}. Relevant sequence parameters for all scans are shown in Table.1.

\begin{table}[!ht]
\renewcommand{\tabcolsep}{0.1cm}
 \caption{\textbf{Scanner and sequence parameters of the phantom and \textit{in vivo} experiments.}}
\begin{center}
           \small{
          \begin{tabular}{l l l l }
            \toprule
           \multicolumn{4}{l}{\textbf{MRF Sequence settings}} \\ 
            & \textbf{Phantom} & \textbf{Brain} & \textbf{Abdomen} \\
            \midrule
            Field strength & 1.5T & 1.5T&1.5T \\
            Spatial resolution & 2.0 x 2.0 mm\textsuperscript{2} & 1.5 x 1.5 mm\textsuperscript{2}  &2.0 x 2.0 mm\textsuperscript{2} \\
            Matrix size & 125 x 125 & 186 x 186 & 175 x 175\\
            Field-of-view & 250 x 250 mm\textsuperscript{2} & 280 x 280 mm\textsuperscript{2} & 350 x 350 mm\textsuperscript{2} \\
            Slick thickness & 10 mm& 5 mm& 10 mm\\
            Repetition time & 5.2 ms & 7.7 ms &5.3 ms \\
            Echo time & 2.5 ms & 3.3 ms & 2.5 ms\\
            Readout bandwidth & 386 Hz/pixel & 285 Hz/pixel & 379 Hz/pixel\\
            N Flip angles & 1000 & 1000 & 1000 \\
            Scan time & 5.2 s & 7.7 s & 5.3 s \\

  \bottomrule
 \end{tabular}
 }
\end{center}
 \label{tbl:1}
\end{table}

\newpage
\section{Results}

\subsection{Phantom studies}
An exemplary time-point image of the fully sampled MRF scan along with the MRF proton density, $T_1$ and $T_2$ parameter maps are shown in Fig.3. The bottom row shows the raw time domain signal (fingerprint) of voxels in tube 1 and 11 along with the match to the dictionary. For both these voxels the time domain signal shows close agreement with the dictionary match. The agreement holds for all the pixels within the tubes with a mean normalized root mean square error (NRMSE) = 0.06. Small differences between the MRF signal and the dictionary match are primarily observed during the first 50 snapshots directly after the inversion pulse and during the higher flip angles in time-points 500-600.  \newline

\begin{figure}[H]
\begin{center}
   \includegraphics[width=16cm]{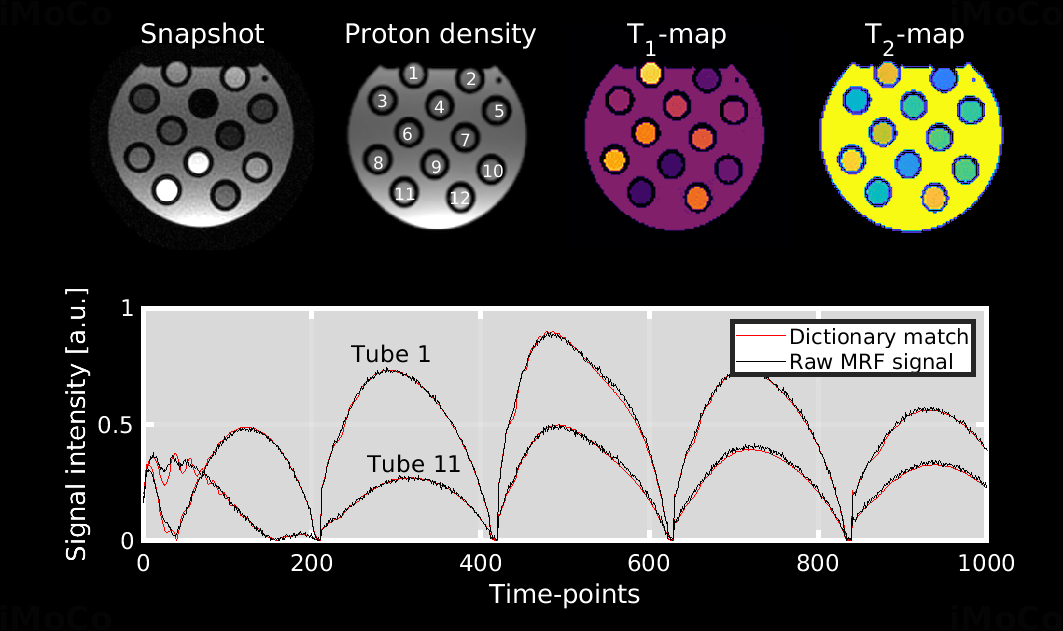}
 \vspace{-5pt}
\caption[]{\label{fig:fig3}\textbf{Analysis of the raw MRF time domain signal.} Top row shows the images reconstructed from the fully sampled MRF measurements. From left-to-right a single time-point image(snapshot) and the reconstructed parameters maps. The bottom row shows the time domain signal of a voxel in tube 1 and a voxel in tube 11. Note that the time signals are in close agreement with the match to the dictionary. See the following link for an animated version: https://surfdrive.surf.nl/files/index.php/s/KavixXHVaQ4c9Ue }\vspace{-5pt}
\end{center}
\end{figure}

\noindent The parameter quantification of the fully sampled (R=1) and undersampled (R=276) MRF reconstructions are compared against the spin-echo reconstructions in Fig.4. Both the R=1 and R=276 MRF reconstructions showed good correlation in average values compared to the spin-echo measurements. The undersampled MRF has coefficients of determination $R_{T_1}^2 = 0.999$ and $R_{T_2}^2 = 0.975$ for $T_1$ and $T_2$, respectively. Note that the accuracy of the $T_{1}$-maps was slightly higher then the $T_{2}$-maps. The precision over all the tubes for the undersampled MRF was $\sigma_{T_1} = 8.6$ ms and $\sigma_{T_2} = 3.0$ ms. 

\begin{figure}[H]
\begin{center}
   \includegraphics[width=16cm]{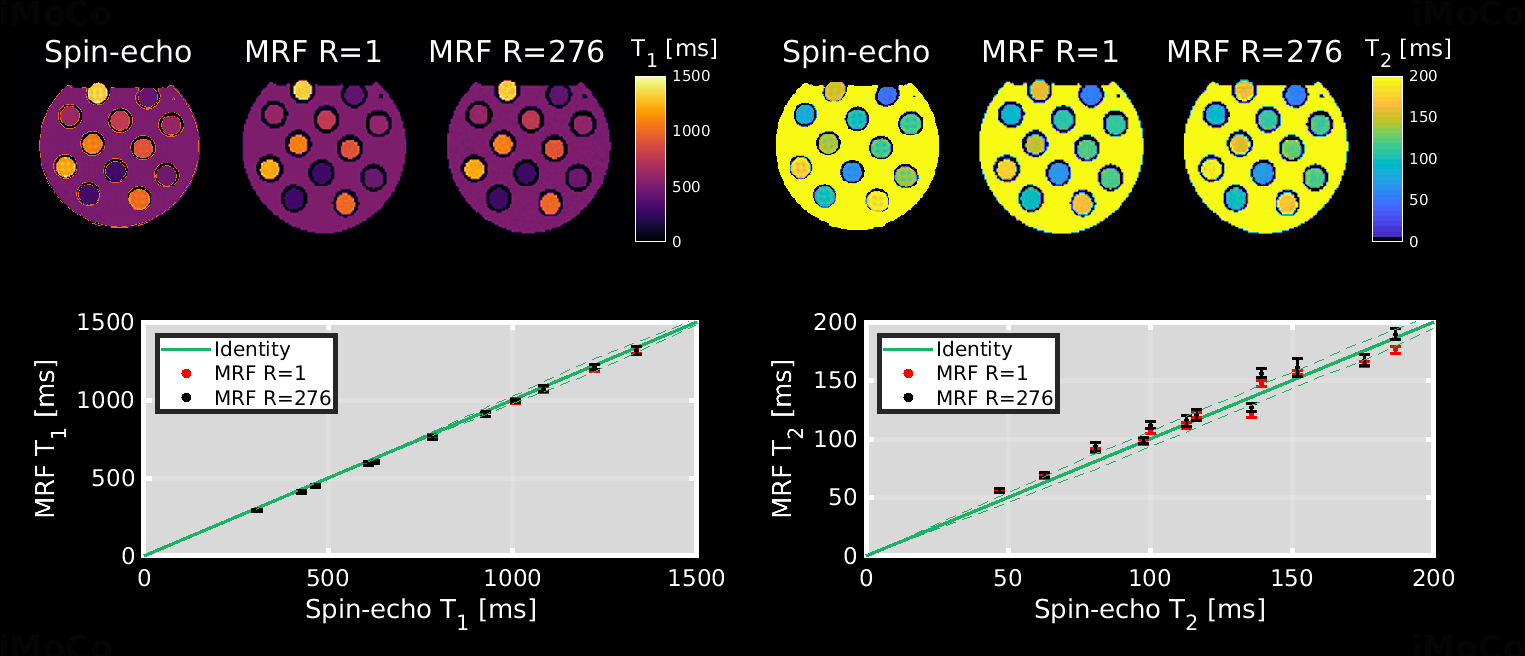}
 \vspace{-5pt}
\caption[]{\label{fig:fig4} \textbf{Accuracy and precision analysis of MRF parameter quantification in a phantom.} Top row shows the spin-echo, fully sampled MRF and undersampled MRF parameter maps. The bottom row shows the correlation of the accuracy estimations for MRF versus the spin-echo. Data show the mean and standard deviation of MRF over a 25 pixel region in the center of the phantom. The dashed green line is the standard deviation of the spin-echo measurements in the same region. }\vspace{-5pt}
\end{center}
\end{figure}

\noindent The temporal stability of the parameter quantification (reproducibility) of the repeated measurements is shown in Fig.5. The $T_1$ values were very stable, while the higher $T_2$ values show slightly higher deviation over time. The mean values within the tubes had an average standard deviation over time of $\sigma_{T_1} = 6.4$ ms and  $\sigma_{T_2} = 2.3$ ms. 

\begin{figure}[H]
\begin{center}
   \includegraphics[width=16cm]{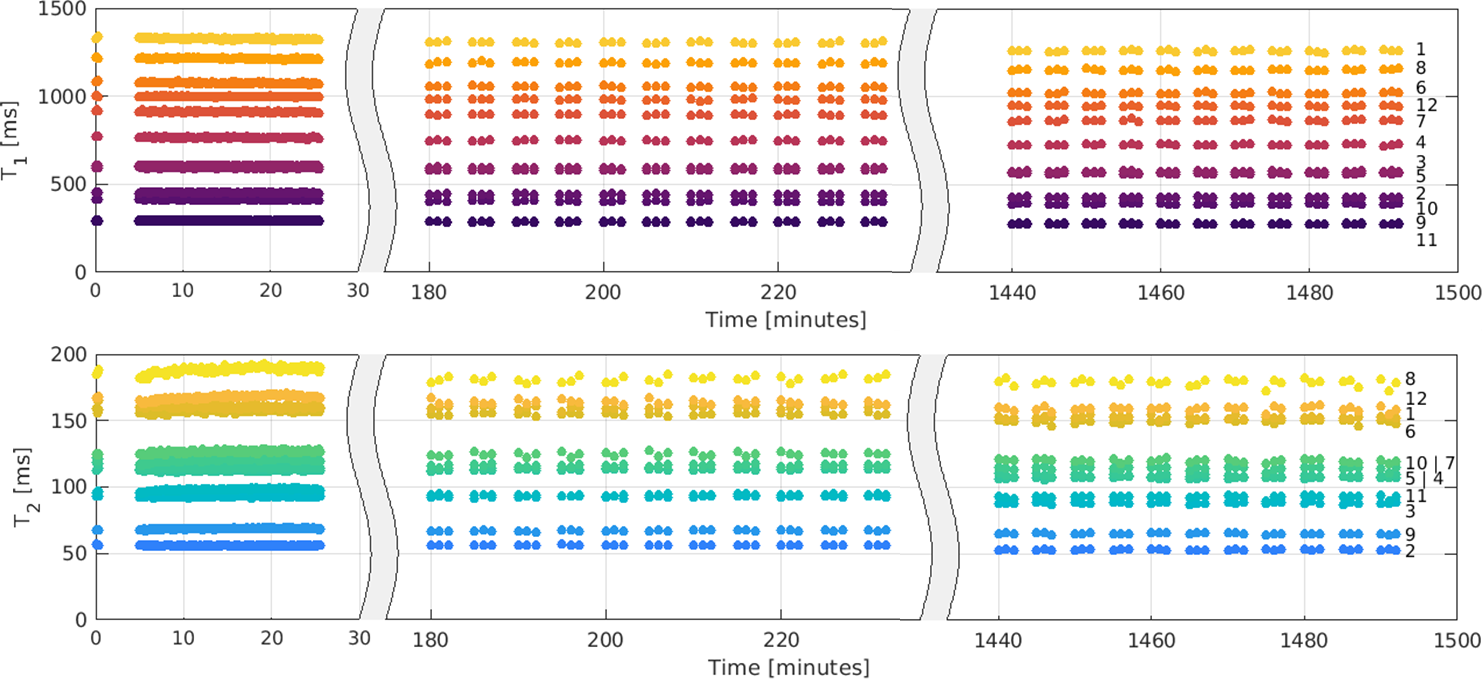}
 \vspace{-5pt}
\caption[]{\label{fig:fig5}\textbf{Repeated MRF measurement to assess the temporal stability of the parameter quantification.} Top row shows the evolution of the mean $T_1$ values within the tubes and the bottom row shows the evolution of the $T_2$ values. Note that the tube numbers are added at the right side of the graph and they correspond with the numbers in Fig.3. See the following link for video that shows the parameter maps over time: https://surfdrive.surf.nl/files/index.php/s/aOB3B2YlAxeT4mo}\vspace{-5pt}
\end{center}
\end{figure}

\subsection{\textit{In-vivo} studies}
\underline{Brain volunteer data}\newline
\noindent Two slices of the brain MRF scans in the volunteer are shown in Fig.6. The $T_1$ and $T_2$ maps show clear boundaries between white and gray matter. The mean parameter values for gray and white matter are within the range of report literature values (Table.2). Note that the $T_2$ values are on the low side, which is also reported in other MRF publications \cite{Jiang2015}. The regions of interest that were used to compute the mean values are shown on the proton density image.

\begin{figure}[H]
\begin{center}
   \includegraphics[width=16cm]{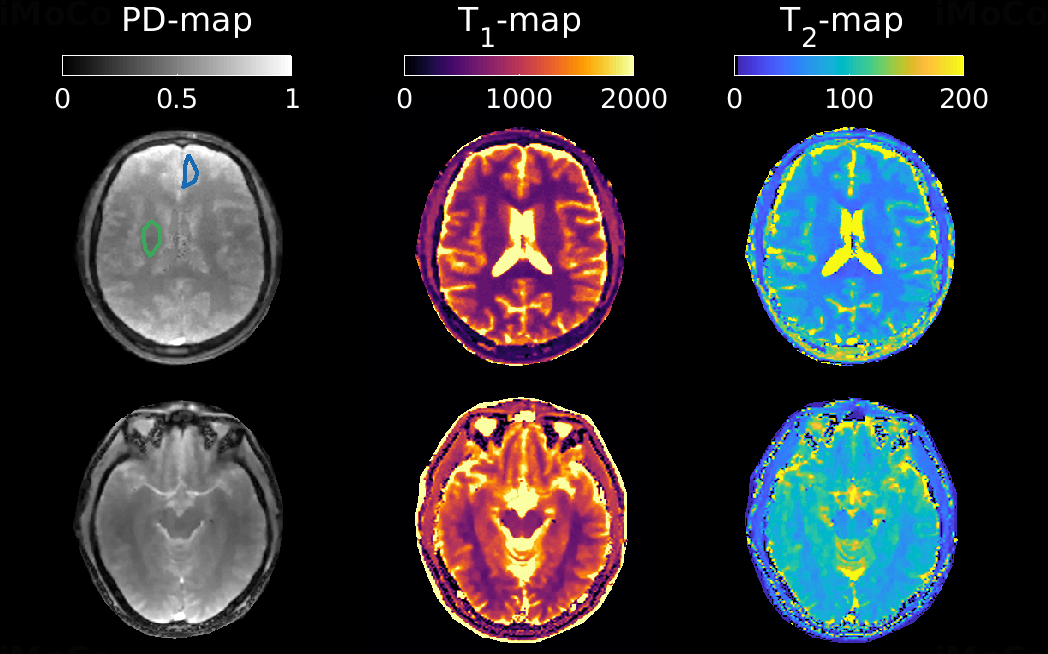}
 \vspace{-5pt}
\caption[]{\label{fig:fig6}\textbf{2D brain MRF measurement in a healthy volunteer.} Top row shows slice 1 of the MRF parameter maps and bottom row shows slice 2. Averaged $T_{1,2}$ values for white matter (green) and gray matter (blue) are shown in Table.2.}\vspace{-5pt}
\end{center}
\end{figure}

\begin{table}[!ht]
\renewcommand{\tabcolsep}{0.2cm}
 \caption{\textbf{Comparison of MRF $T_1$ and $T_2$ quantification to literature reported values. cite(de Bazelaire) }}
\begin{center}
           \small{
          \begin{tabular}{l l l l l}
            \toprule
           \multicolumn{5}{l}{\textbf{MRF parameter quantification}} \\ 
            & \textbf{Reference $T_1$} & \textbf{MRF $T_1$} & \textbf{Reference $T_2$} & \textbf{MRF $T_2$} \\
            \midrule
            White matter & 608 - 756 ms & 626 $\pm$ 34  ms & 54 - 81 ms& 56 $\pm$ 4 ms\\
            Gray matter & 998 - 1304 ms& 1113 $\pm$ 91 ms & 78 - 98 ms&76 $\pm$ 7 ms \\
            Liver & 547 - 625 ms& 612 $\pm$ 42 ms & 40 - 52 ms&46 $\pm$ 5 ms \\
            Kidney (medulla) & 1354 - 1470 ms& 1510 $\pm$ 144 ms & 74 - 96 ms&51 $\pm$ 6 ms \\
            Kidney (cortex) & 908 - 1024 ms& 954 $\pm$ 85 ms & 83 - 91 ms&54 $\pm$ 5 ms \\
            Pancreas & 570 - 598 ms& 540 $\pm$ 59 ms & 40 - 52 ms&47 $\pm$ 8 ms \\
  \bottomrule
 \end{tabular}
 }
\end{center}
 \label{tbl:2}
\end{table}
\noindent\underline{Volunteer abdomen data}\newline
\noindent Two slices of the abdomen MRF scans in the volunteer are shown in Fig.7-8. The boundaries between the medulla and cortex of the kidney are well defined on both the $T_1$-map and the $T_1$-\textit{w} image, while the boundary is not visible on the $T_2$-\textit{w} image and $T_2$-map. On the left side of the liver a small benign lesion is clearly visible on both the $T_1$ and $T_2$ map, which is characterised with a high $T_1$ and high $T_2$. The $T_2$ values differ between the right and left kidney, which are 35 and 51 ms respectively. Region of interest analysis for multiple organs are shown in Table.2. The kidney $T_2$ values also differ slightly between the two scans. However, the $T_1$ values were constant between the left and right kidney and between slice 1 and 2. The regions of interest that were used to compute the mean values are shown on the proton density image.

\begin{figure}[H]
\begin{center}
   \includegraphics[width=16cm]{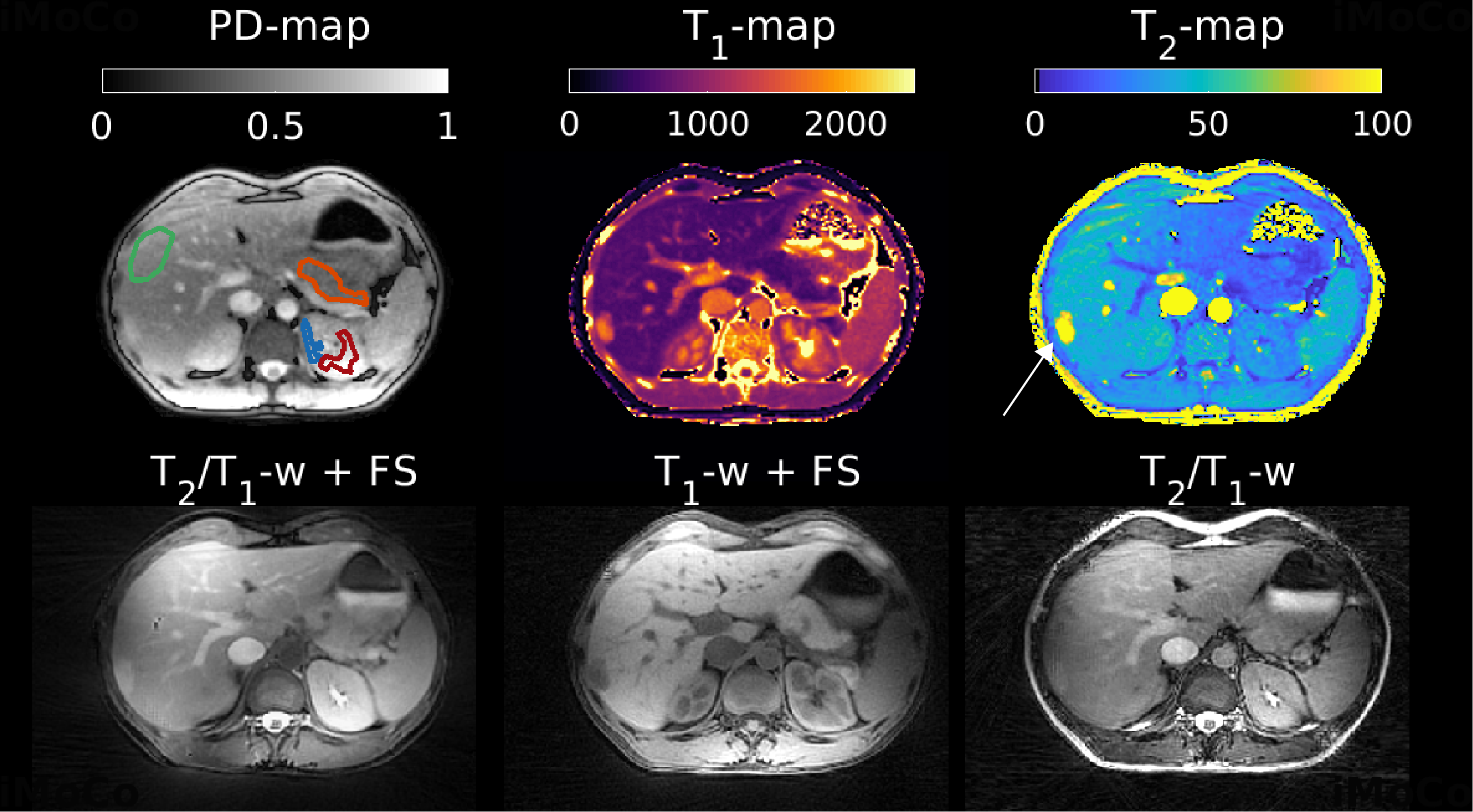}
 \vspace{-5pt}
\caption[]{\label{fig:fig7}\textbf{Breathhold 2D MRF measurement in a healthy volunteer.} Top row shows the MRF parameter maps. Bottom row shows the qualitative images derived a clinical protocol. The $T_1$-\textit{w} scan is a spoiled gradient echo sequence, the $T_2/T_1$-\textit{w} is a balanced gradient echo scan and FS = fat suppression. Note that the lesion that is visible in the liver is a benign cyst indicated by the white arrow. Mean $T_1$ and $T_2$ values were analyzed in regions of interest for liver (green), pancreas (orange), medulla (red) and cortex (blue) of the kidney.}\vspace{-5pt}
\end{center}
\end{figure}

\begin{figure}[H]
\begin{center}
   \includegraphics[width=16cm]{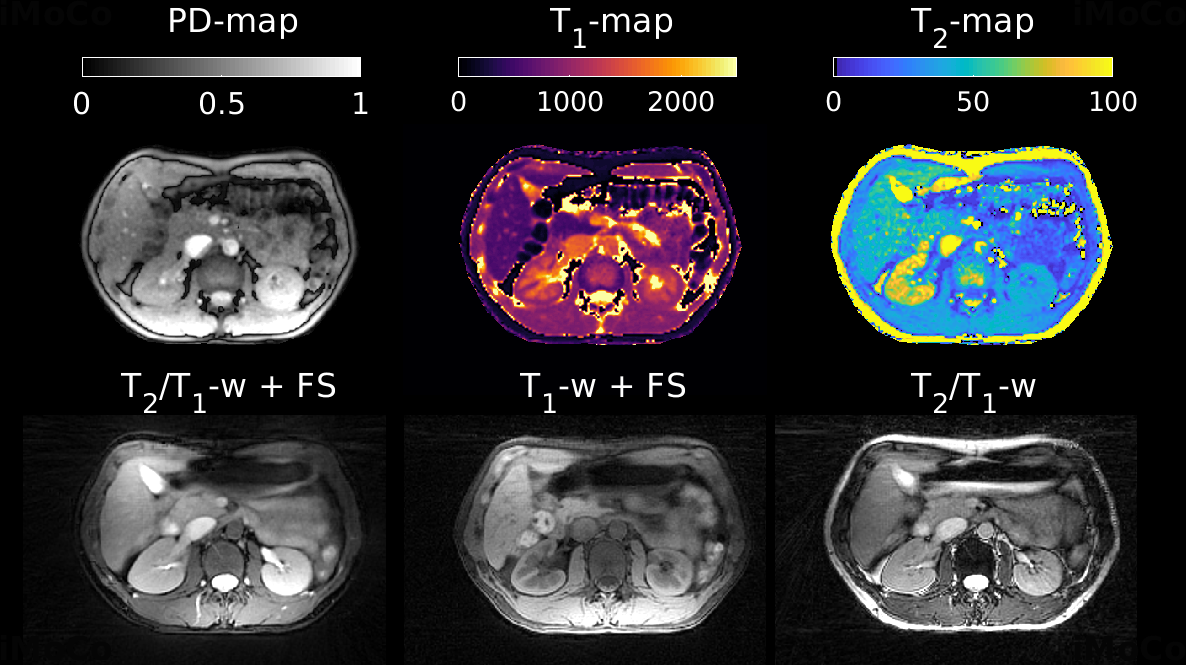}
 \vspace{-5pt}
\caption[]{\label{fig:fig8}\textbf{Breathhold 2D MRF measurement in a healthy volunteer.} Top row shows the MRF parameter maps. Bottom row shows the qualitative images derived a clinical protocol. The $T_1$-\textit{w} scan is a spoiled gradient echo sequence, the $T_2/T_1$-\textit{w} is a balanced gradient echo scan and FS = fat suppression.}\vspace{-5pt}
\end{center}
\end{figure}

\noindent\underline{Patient abdomen data} \newline
\noindent One slice of the abdomen MRF scan in the patient with a recurrent colorectal liver metastasis after hepatic resection is shown in Fig.9. The metastasis is positioned in the anterior side of the liver and is clearly visible on the $T_1$-map, $T_1$-\textit{w} image and on the diffusion-\textit{w} image, while the lesion is less well defined on the $T_2$-map and $T_2$-\textit{w} image. The $T_2$-map shows lower values in the liver and spleen compared to the volunteer scans, which could be due to patient motion.
\begin{figure}[H]
\begin{center}
   \includegraphics[width=16cm]{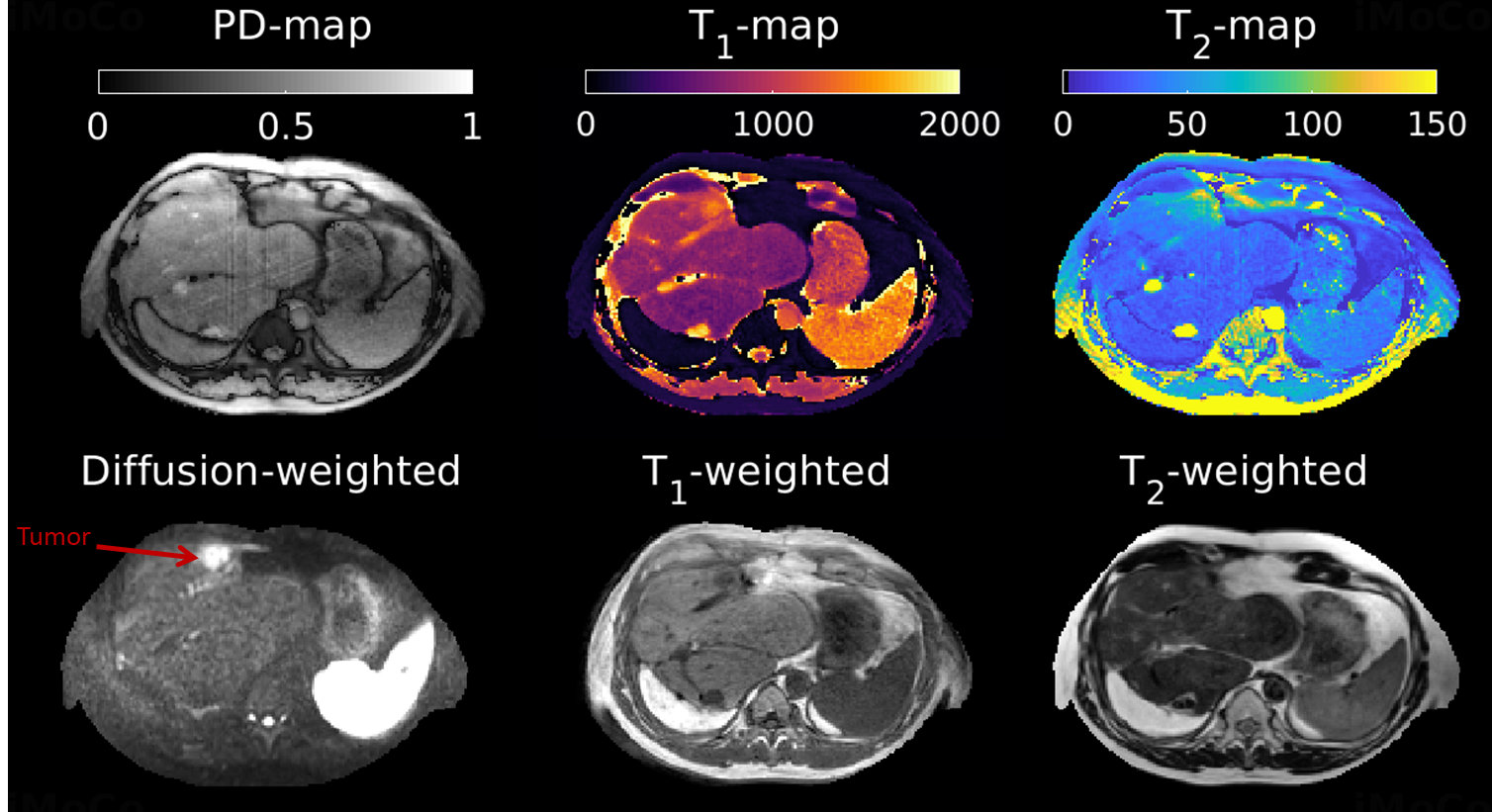}
 \vspace{-5pt}
\caption[]{\label{fig:fig9}\textbf{Free-breathing 2D MRF measurement in a patient with a colorectal liver metastasis.} Top row shows the MRF parameter maps. Bottom row shows the qualitative images from the clinically used protocol. The $T_1$-\textit{w} scan is a spoiled gradient echo sequence, the $T_2$-\textit{w} is a turbo spin echo sequence and the diffusion-\textit{w} is a spin-echo sequence with EPI readout. The lesion is indicated with the red arrow on the diffusion-weighted image and is also clearly visible on the $T_1$ scans. }\vspace{-5pt}
\end{center}
\end{figure}

\newpage
\section{Discussion}
In this study we demonstrated technical feasibility of 2D MRF on a 1.5T MRL system. The phantom study indicated good agreement of parameter quantification ($R_{T_1}^2=0.999$ and $R_{T_2}^2=0.975$) with reference measurements, high precision ($\sigma_{T_1}=8.6 ms$ and $\sigma_{T_2}=3.0 ms$) and temporally stable measurements during the day ($\sigma_{T_1}=6.4 ms$ and $\sigma_{T_2}=2.3 ms$). The \textit{in vivo} study showed high image quality of the fast MRF scans, where image features in the quantitative maps nicely corresponded with the qualitative scans. We believe these observations provide sufficient evidence that MRF is technically feasible on MRL systems and therefore could be further explored for online MR-guided radiotherapy applications on a 1.5T MRL. Besides MRF, these findings also apply for other transient-state parameter quantification methods \cite{Sbrizzi2018}.\newline

\noindent A possible use case for the quantitative maps could be patient-specific contrast optimization of the anatomical turbo spin-echo sequences, i.e. the reference MRI on which the treatment is planned. For example, liver metastasis are a heterogeneous group of lesions  that  show  variable  signal  characteristics on both $T_1w$ and $T_2w$ imaging depending on the primary origin \cite{Danet2003,Namasivayam2007}. In this context, MRF could function as a contrast scout scan followed by an on-the-fly flip angle train optimization to maximize the contrast-to-noise ratio between the lesion and the liver. Contrast optimization techniques are well described in literature \cite{Sbrizzi2017}, but have never been applied in an on-the-fly setting for online contrast optimization on either diagnostic MR systems or MRL systems. Future work will focus on the implementation of these patient-specific contrast optimization techniques to investigate the potential improvement in image quality. \newline

\noindent The rapid acquisition scheme of MRF ($\approx$ 5 s per slice) could facilitate the integration of quantitative imaging to the clinical MRI-guided radiotherapy workflow without significantly lengthening of the treatment. The primary application of MRF would be for tumor response monitoring over multiple fractions during the treatment. The optimal timing to image changes in quantitative parameters post radiotherapy is an active topic of research \cite{Fang2018,VanSchie2019,Borggreve2019} and could be pushed forward with daily MRF on the MRL. The ability to pick up subtle changes in $T_1$ and $T_2$ values could be used to distinguish responders from non-responders. Ultimately, these potential changes in $T_1$ and $T_2$ could be used to intensify or reduce the (local) radiation during the radiotherapy treatment period based on the measured response.\newline

\section{Conclusion}
Gradient spoiled 2D magnetic resonance fingerprinting is feasible on a 1.5T MRI-Linac with similar performance as on a diagnostic system. The precision and accuracy of the parametric maps are sufficient for further investigation of the clinical utility of magnetic resonance fingerprinting for online quantitatively MRI-guided radiotherapy.
\label{sec:Concl}

%\ack % or \ackn
\section*{Acknowledgements}
\addcontentsline{toc}{section}{Acknowledgements}
This work is part of the research programme HTSM with project number 15354, which is (partly) financed by the Netherlands Organisation for Scientific Research (NWO) and Philips Healthcare.

%\setcounter{section}{1}
%\appendix

\section*{References}
\addcontentsline{toc}{section}{References}
\bibliography{main}

\newpage
\begin{comment}
\section{Supplementary Material}
\end{comment}
\end{document}